\documentclass[aps,preprint,prd,nofootinbib]{revtex4}

\usepackage{graphicx}
\usepackage{psfrag}

\usepackage{subfigure}
\usepackage{array}
\usepackage{graphicx}
\usepackage{amsmath}
\usepackage{amssymb}
\usepackage{mathrsfs}
\usepackage{bm}
\usepackage{slashed}
\usepackage{threeparttable}
\usepackage{multirow}

\usepackage[colorlinks=true, linkcolor=blue, citecolor=blue, urlcolor=blue]{hyperref}

\usepackage[amsmath,thmmarks,hyperref]{ntheorem}
\usepackage{soul}

\allowdisplaybreaks[4]

\begin{document}

\title{Revisiting the Heavy Vector Quarkonium Leptonic Widths}
\author{Guo-Li Wang$^1$}
\email{gl\_wang@hit.edu.cn}
\author{Xing-Gang Wu$^2$}
\email{wuxg@cqu.edu.cn}
\address{1, Department of Physics, Hebei University, Baoding 071002, China\\
2, Department of Physics, Chongqing University, Chongqing 401331, China}

\baselineskip=20pt

\begin{abstract}

We revisit the heavy quarkonium leptonic decays $\psi(nS) \to \ell^+\ell^-$ and $\Upsilon(nS) \to \ell^+\ell^-$ using the Bethe-Salpeter method. The emphasis is on the relativistic correction. For the $\psi(1S-5S)$ decays, the relativistic effects are $22^{+3}_{-2}\%$, $34^{+5}_{-5}\%$, $41^{+6}_{-6}\%$, $52^{+11}_{-13}\%$ and $62^{+14}_{-12}\%$, respectively. For the $\Upsilon(1S-5S)$ decays, the relativistic effects are $14^{+1}_{-2}\%$, $23^{+0}_{-3}\%$, $20^{+8}_{-2}\%$, $21^{+6}_{-7}\%$ and $28^{+2}_{-7}\%$, respectively. Thus, the relativistic corrections are large and important in heavy quarkonium leptonic decays, especially for the highly excited charmonium. Our results for $\Upsilon(nS) \to \ell^+\ell^-$ are consistent with the experimental data.

\end{abstract}

\maketitle

\section{Introduction}

As it gives a clean experimental signal, the dilepton annihilation decay of the heavy vector quarkonium plays an important role in determining the fundamental parameters such as the strong coupling constant \cite{penin0, beneke5}, heavy quark masses \cite{penin0, penin, beneke3, beneke4}, heavy quarkonium decay constants \cite{beneke5, badalian, gonzalez, chao}, etc. Its decay amplitude is a function of the quarkonium wave function, and this process can be used to test various theories such as the quark potential model, non-relativistic Quantum Chromodynamics (NRQCD), QCD sums rules,  lattice QCD, etc. The Standard Model prediction of the universality of lepton flavor is questioned by the measured ratios $R(D^{(*)})$ and $R(K^{(*)})$~\cite{sanchez, cinabro, CLEO, besson, aloni, lozano}, and the quarkonium leptonic decay is another way to test the lepton flavor universality.

The vector quarkonium leptonic decays have been studied since a long time \cite{barbieri, bergstrom, niczyporuk, eichten, eichten1, buchmuller, quigg}. With the progress in computer science and experimental technology, many advances have been reported in  literature. For example, on can find lattice QCD predictions of the leptonic decays of the ground-state $\Upsilon$ and its first radial excitation $\Upsilon'$ in \cite{colquhoun}; Ref. \cite{pineda} reported the next-to-leading non-perturbative prediction and Ref. \cite{pineda1} the next-to-leading-log perturbative QCD (pQCD) prediction; in Ref. \cite{beneke}, the two-loop QCD correction was computed; Ref. \cite{pineda2} studied the inclusive leptonic decay of $\Upsilon$ up to the next-to-next-to-leading order (NNLO) by including the re-summation of the logarithms (partly) up to the next-to-next-to-leading logarithmic (NNLL) accuracy; the NNNLO corrections have been discussed by various groups \cite{beneke1, kniehl, marquard1, beneke2, marquard}. A pQCD analysis of the $\Upsilon(1S)$ leptonic decay up to NNNLO using the principle of maximum conformality (PMC) \cite{Brodsky:2011ta, Brodsky:2012rj, Mojaza:2012mf, Brodsky:2013vpa} was presented in Refs.\cite{wu, wu2}, where the renormalization scale ambiguity of the decay width is eliminated with the help of the renormalization group equation.

Even though considerable improvements have been made, there are still deviations between the theoretical predictions and the experimental data for the heavy vector quarkonium leptonic decays. There are two sources which may cause such deviations. The first are the unknown higher order perturbative QCD corrections. By using  PMC, the conventional pQCD convergence of the series can be greatly improved by the elimination of the divergent renormalon terms, and a more accurate decay width can be obtained. However, there are still large errors due to unknown high-order terms \cite{wu, wu2}. The second source is the relativistic correction, which could be large. However, almost all pQCD predictions are calculated using NRQCD, in which the decay constant of quarkonium, or its wave function at the origin is treated simply in the non-relativistic approximation.

One may argue that the relativistic correction is small for a heavy quarkonium, since the relative velocity among the heavy constituent quarks is small, e.g. ${v}^2 \sim 0.1-0.3$~\cite{nrqcd}. However, many analyses in literature have found that the relativistic effect could be large. For example, Bodwin {\it et al.} computed the coefficients of the decay operators for the $^3 S_1$ heavy quarkonium decay into a leptonic pair and found large relativistic correction \cite{bodwin}; Gonzalez {\it et al.} pointed out that large relativistic and QCD corrections of the  quarkonium leptonic decays are necessary to fit the experimental data \cite{gonzalez}; Geng {\it et al.} studied the $B_c$ meson semileptonic decays into charmonium and also found that the relativistic corrections are  large \cite{geng}, especially for highly excited charmonium states. Moreover, from the experimental standpoint, Ref. \cite{yuan} showed that a careful study of leptonic decays is still needed for highly excited charmonium states.

In this paper, we focus on the leptonic decays of charmonium and bottomonium, including their excited states, using the relativistic method. In a previous short letter \cite{fu}, we presented a relativistic calculation of the quarkonium decays into $e^{+}e^{-}$, where the results disagree with the experimental data. As a step forward, we revisit this topic in more detail, and include the decays into $\mu^{+}\mu^{-}$ and $\tau^{+}\tau^{-}$ as well as the ratios $R_{\tau\tau}$. We present the relativistic effects in these quarkonium decays, and discuss the universality of lepton flavor.

The paper is organized as follows. The general equation of the quarkonium leptonic decay width is given in Sec.2. In Sec.3, we give a brief review of the Bethe-Salpeter equation, and its instantaneous version, the Salpeter equation. We then show in Sec.4 in detail how to solve the full Salpeter equation and obtain the relativistic wave function for a vector meson. The calculation of the decay constant in the relativistic method is given in Sec.5. Finally, in Sec.6, we give the numerical results and a discussion. A summary is presented in Sec.7.

\section{The quarkonium leptonic Decay Width }

The leptonic partial decay rate of a vector charmonium or bottomonium $nS$ state $V$ is given by
\begin{equation}
\Gamma_{V\to \ell^{+}\ell^{-}}=\frac{4\pi\alpha^2_{em}e_Q^2F^2_V}{3M_{nS}}\times
\left(1+2\frac{m^2_{\ell}}{M^2_{nS}}\right)\sqrt{1-4\frac{m^2_{\ell}}{M^2_{nS}}},
\end{equation}
where $\alpha_{em}$ is the fine structure constant, $e_Q$ is the electric charge of the heavy quark $Q$ in units of the electron charge, $e_Q=+2/3$ for the charm quark and $e_Q=-1/3$ for the bottom quark, $M_{nS}$ is the mass of the $nS$ state quarkonium, $m_{\ell}$ is the lepton mass, $F_V$ is the decay constant of the vector meson that is defined by the following matrix element of the electromagnetic current
\begin{equation}
<0|\bar Q\gamma_{\mu} Q|V(P,\epsilon)>=F_V M_{nS}\epsilon_{\mu},\label{decayconstant}
\end{equation}
where $P$ is the quarkonium momentum, and $\epsilon$ is the polarization vector.

In the non-relativistic method, the well-known formula for the decay constant is
\begin{equation}
F^{NR}_V=\sqrt{\frac{12}{M_{nS}}}|\Psi_V(0)|,
\end{equation}
where $NR$ means the non-relativistic (NR), and $\Psi_V(0)$ is the non-relativistic wave function evaluated at the origin. In the NR method, there is one radial wave function, and the vector meson and its corresponding pseudoscalar have the same radial wave function and the same decay constant. However, in the relativistic method, they have different wave functions and different decay constants, and more than one radial wave function gives a contribution to the vector meson decay constant.

In the relativistic method, the decay constant $F_V=F_V^{Re}$ is not related to the wave function at the origin, but in the full region. In the following, we focus on the calculation of $F_V^{Re}$ in the relativistic method.

\section{The Bethe-Salpeter Equation and the Salpeter Equation}

In this section, we briefly review the Bethe-Salpeter (BS) equation \cite{BS}, which is a relativistic dynamic equation
describing the two-body bound state, and its instantaneous version, the Salpeter equation \cite{salp}. The BS equation for a meson, which is a bound state of a quark, labelled as 1, and anti-quark, labelled as 2, can be written as \cite{BS}
\begin{equation}
(\slashed{p}_{1}-m_{1})\chi_{_P}(q)(\slashed{p}_{2}+m_{2})= i\int\frac{d^{4}k}{(2\pi)^{4}}V(P,k,q)\chi_{_P}(k)\;, \label{eq2-1}
\end{equation}
where $\chi_{_P}(q)$ is the relativistic wave function of the meson, $V(P,k,q)$ is the interaction kernel between the quark and anti-quark, $p_{1}, p_{2},m_1,m_2$ are the momenta and masses of the quark and anti-quark, $P$ is the momentum of the meson, $q$ is the relative momentum between quark and anti-quark. The momenta $p_1$ and $p_2$ satisfy the relations, $p_{1}={\alpha}_{1}P+q$ and $p_{2}={\alpha}_{2}P-q$, where ${\alpha}_{1}=\frac{m_{1}}{m_{1}+m_{2}}$ and ${\alpha}_{2}=\frac{m_{2}}{m_{1}+m_{2}}$. In the case of quarkonium, where $m_1=m_2$, we have ${\alpha}_{1}={\alpha}_{2}=0.5$.

In the general case, the BS equation is hard solve due to the complex interaction kernel between the constituent quarks. For the doubly heavy quarkonium considered here, the interaction kernel between the two heavy constituent quarks can be treated as instantaneous, leading to a simpler version of the BS equation. In this case, it is convenient to divide the relative momentum $q$ into two parts, $q^{\mu}=q^{\mu}_{\parallel}+q^{\mu}_{\perp}$, where $q^{\mu}_{\parallel}\equiv (P\cdot q/M^{2})P^{\mu}$ and
$q^{\mu}_{\perp}\equiv q^{\mu}-q^{\mu}_{\parallel}$, $M$ is the mass of the bound state, and we have $P^2=M^2$. Then we have two Lorentz invariant variables, $q_{_P}=\frac{(P\cdot q)}{M}$ and $q_{_T}=\sqrt{q^{2}_{_P}-q^{2}}=\sqrt{-q^{2}_{\perp}}$. When $\vec{P} =0$, that is in the meson center-of-mass frame, they reduce to the usual components $q_{0}$ and $|\vec q|$, and $q_{\perp}=(0,\vec{q})$.

With this notation, the volume element of the relativistic momentum $k$ can be written in an invariant form $d^4k=dk_{_P} k^2_{_T} {{dk_{_T}}} ds d\phi$, where $ds =(k_{_P} q_{_P} -k\cdot q)/(k_{_T} q_{_T})$ and $\phi$ is the azimuthal angle. Taking the instantaneous approximation in the center-of-mass frame of the bound state, the kernel $V(P,k,q)$ changes to $ V(k_{\perp},q_{\perp},s)$. We introduce the three-dimensional wave function
\begin{equation}
\Psi_{P}(q^{\mu}_{\perp})\equiv i\int
\frac{dq_{_P}}{2\pi}\chi(q^{\mu}_{\parallel},q^{\mu}_{\perp}),
\end{equation}
and the notation
\begin{equation}
\eta(q^{\mu}_{\perp})\equiv\int\frac{k^2_{_T}dk_{_T} ds {{d \phi}}}{(2\pi)^{{{3}}}}
V(k_{\perp},q_{\perp},s)\Psi_{P}(k^{\mu}_{\perp}).
\end{equation}
The BS equation Eq. (\ref{eq2-1}) is then rewritten as
\begin{equation}
\chi(q_{\parallel},q_{\perp})=S_{1}(p_{1})\eta(q_{\perp})S_{2}(p_{2}),
\label{eq2-2}
\end{equation}
where $S_{1}(p_{1})$ and $S_{2}(p_{2})$ are propagators of quark 1 and anti-quark 2, respectively, which can be decomposed as
\begin{equation}
S_{i}(p_{i})=\frac{\Lambda^{+}_{i}(q_{\perp})}{(-1)^{i+1} q_{_P} +\alpha_{i}M-\omega_{i}+i\varepsilon}+
\frac{\Lambda^{-}_{i}(q_{\perp})}{(-1)^{i+1} q_{_P}+\alpha_{i}M+\omega_{i}-i\varepsilon}\;.
\label{eq2-3}
\end{equation}
Here, we have defined the constituent quark energy $\omega_{i}=\sqrt{m_{i}^{2}+q^{2}_{_T}}$ and the projection operators
$\Lambda^{\pm}_{i}(q_{\perp})= \frac{1}{2\omega_{i}}\left[ \frac{\slashed{P}}{M}\omega_{i}\pm
(-1)^{i+1} (m_{i}+\slashed{q}_{\perp})\right]$, where $i=1$ and $2$ for quark and anti-quark, respectively.

Using the projection operators, we can divide the wave function into four parts
\begin{equation}
\Psi_{P}(q_{\perp})=\Psi^{++}_{P}(q_{\perp})+
\Psi^{+-}_{P}(q_{\perp})+\Psi^{-+}_{P}(q_{\perp})
+\Psi^{--}_{P}(q_{\perp}), \label{eq2-4}
\end{equation}
with the definition $\Psi^{\pm\pm}_{P}(q_{\perp})\equiv \Lambda^{\pm}_{1}(q_{\perp})
\frac{\slashed{P}}{M}\Psi_{P}(q_{\perp}) \frac{\slashed{P}}{M} \Lambda^{{\pm}}_{2}(q_{\perp})$. Here $\Psi^{++}_{P}(q_{\perp})$ and $\Psi^{--}_{P}(q_{\perp})$ are called the positive and negative energy wave functions of the quarkonium.

After integrating over $q_{_P}$ on both sides of Eq. (\ref{eq2-2}) using contour integration, we obtain the famous Salpeter equation \cite{salp}:
\begin{equation}
\Psi_{P}(q_{\perp})=\frac{
\Lambda^{+}_{1}(q_{\perp})\eta(q_{\perp})\Lambda^{+}_{2}(q_{\perp})}
{(M-\omega_{1}-\omega_{2})}- \frac{
\Lambda^{-}_{1}(q_{\perp})\eta(q_{\perp})\Lambda^{-}_{2}(q_{\perp})}
{(M+\omega_{1}+\omega_{2})}.\label{salpe}
\end{equation}
Equivalently, the Salpeter equation can be written as four independent equations using the projection operators:
\begin{equation}
(M-\omega_{1}-\omega_{2})\Psi^{++}_{P}(q_{\perp})=
\Lambda^{+}_{1}(q_{\perp})\eta(q_{\perp})\Lambda^{+}_{2}(q_{\perp})\;,\label{positive}
\end{equation}
\begin{equation}
(M+\omega_{1}+\omega_{2})\Psi^{--}_{P}(q_{\perp})=-
\Lambda^{-}_{1}(q_{\perp})\eta(q_{\perp})\Lambda^{-}_{2}(q_{\perp})\;,\label{negative}
\end{equation}
\begin{equation}
\Psi^{+-}_{P}(q_{\perp})=0\;,
\label{pone}
\end{equation}
\begin{equation}
\Psi^{-+}_{P}(q_{\perp})=0\;.
\label{nepo}
\end{equation}

The normalization condition for the BS wave function reads
\begin{equation}
\int\frac{q_{_T}^2dq_{_T}}{2{\pi}^2}Tr\left[\overline\Psi^{++}_{P}
\frac{{/}\!\!\!
{P}}{M}\Psi^{++}_{P}\frac{{/}\!\!\!{P}}{M}-\overline\Psi^{--}_{P}
\frac{{/}\!\!\! {P}}{M}\Psi^{--}_{P}\frac{{/}\!\!\!
{P}}{M}\right]=2M\;. \label{nor}
\end{equation}

Note that usually in literature, it is not the full Salpeter equation Eq. (\ref{salpe}) that is solved (or equivalently, the four Eqs. (11-14)), but only Eq. (\ref{positive}), which involves only the positive wave function. There is a good reason why such an approximation is made: it is its effective range. The numerical value of $M-\omega_{1}-\omega_{2}$ in Eq. (\ref{positive}) is much smaller than of $M+\omega_{1}+\omega_{2}$ in Eq. (\ref{negative}), which means that the positive wave function $\Psi^{++}_P(q_{\perp})$ is dominant, and that the contribution of the negative wave function $\Psi^{--}_P(q_{\perp})$ can be safely neglected. However, we point out that if only Eq. (\ref{positive}) for $\Psi^{++}_P(q_{\perp})$ is considered, then not only is the contribution of the negative wave function neglected, but so are the relativistic effects of these wave functions. The reason is that the number of eigenvalue equations limits the number of radial wave functions, and as is shown below, only the four coupled equations Eqs. (11-14) can provide sufficient information to derive a relativistic wave function.

\section{Relativistic Wave Function and the Kernel}

Although BS or the Salpeter equation is the relativistic dynamic equation describing the two-body bound state, the equation cannot by itself provide the information about the wave function. This means that we need to provide an explicit form of the relativistic kinematic wave function as input, which can be constructed using all allowable Lorentz and $\gamma$ structures.

From literature, we have the familiar form of the non-relativistic wave function for the $1^-$ vector meson, e.g.
\begin{equation}
\Psi_P(\vec{q})=(\slashed P+M)\slashed{\epsilon} \psi(\vec{q}), \label{eq:schrodinger}
\end{equation}
where $M$, $P$ and $\epsilon$ are the mass, momentum and polarization of the vector meson, $\vec{q}$ is the relative momentum between the quark and anti-quark. There is only one unknown wave function $\psi(\vec{q})$ in Eq.(\ref{eq:schrodinger}), which can be obtained numerically by solving Eq. (\ref{positive}) or the non-relativistic Schr\"odinger equation. The relative momentum $\vec{q}$ is related to the relative velocity $\vec v$ between the quark and anti-quark in the meson, $\vec q = \frac{m_1 m_2}{m_1+m_2}\vec {v}$. A relativistic wave function should depend on the relative velocity $\vec {v}$ or momentum $\vec{q}$ separately, not merely on the radial part $\psi(\vec {q})$, because the radial part is in fact $\psi(|\vec {q}|)$ or equally $\psi(\vec {q}^2)$.

To ontain the form of the relativistic wave function, we start from $J^{pc}$ of a meson, because $J^{p}$ or $J^{pc}$ are in any case good quantum numbers, where $J$ is the total angular momentum, and $p$ and $c$ are the parity and the charge conjugate parity of the meson. The parity transform changes the momentum $q=(q_0,\vec{q})$ into $q'=(q_0,-\vec{q})$, so for a meson, after applying the parity transform, the four-dimensional wave function $\chi_{_P}(q)$ changes to $p\cdot\gamma_0\chi_{_{P'}}(q')\gamma_0$, where $p$ is the eigenvalue of parity. The charge conjugate transform changes $\chi_{_P}(q)$ to $c\cdot
\mathcal{C}\chi^{T}_{_P}(-q)\mathcal{C}^{-1}$, where $c$ is the eigenvalue of charge conjugate parity, $\mathcal{C}=\gamma_2\gamma_0$ is the charge conjugate transform operator, $T$ is the transpose transform. Since the Salpeter equation is instantaneous, the input wave function
$\Psi_{_P}(q_{\perp})$ is also instantaneous, and the general form of the wave function for the $1^{-}$ vector meson can be written as
\cite{chen1, wang1}
\begin{eqnarray}
\Psi_{_P}^{1^{-}}(q_{\perp}) &=& q_{\perp}\cdot{\epsilon}_{\perp}
\left[\psi_1(q_{\perp})+\frac{\not\!P}{M}\psi_2(q_{\perp})+
\frac{{\not\!q}_{\perp}}{M}\psi_3(q_{\perp})+\frac{{\not\!P}
{\not\!q}_{\perp}}{M^2} \psi_4(q_{\perp})\right]+
M{\not\!\epsilon}_{\perp}\psi_5(q_{\perp}) \nonumber\\
&&+ {\not\!\epsilon}_{\perp}{\not\!P}\psi_6(q_{\perp})+
({\not\!q}_{\perp}{\not\!\epsilon}_{\perp}-
q_{\perp}\cdot{\epsilon}_{\perp})
\psi_7(q_{\perp})+\frac{1}{M}({\not\!P}{\not\!\epsilon}_{\perp}
{\not\!q}_{\perp}-{\not\!P}q_{\perp}\cdot{\epsilon}_{\perp})
\psi_8(q_{\perp}).   \label{eq3-2}
\end{eqnarray}
There are in total $8$ radial wave functions $\psi_i(q_{\perp})=\psi_i(|\vec q|)$ with $i=1\sim8$, which obviously can not be obtained by solving only one equation, e.g. Eq. (\ref{positive}), but can be obtained by solving the full Salpeter  Eqs. (11-14). The above expression does not include the terms with $P\cdot q$, since the condition of instantaneous interaction is $P\cdot q =P \cdot q_{\perp}=0$. There are also no higher order $q_{\perp}$ terms like $q^2_{\perp}$, $q^3_{\perp}$, $q^4_{\perp}$, etc., because the even powers of $q_{\perp}$ can be absorbed into the radial part of $\psi_i(q_{\perp})$, while the odd powers of $q_{\perp}$ can be changed to lower power, for example, ${\not\!q}^3_{\perp}\psi'_i(q_{\perp}) ={\not\!q}_{\perp}\psi_i(q_{\perp})$. By the way, if we delete all $q_\perp$ terms except those inside the radial wave functions, then the wave function Eq. (\ref{eq3-2}) reduces to
$M{\not\!\epsilon}_{\perp}\psi_5(q_{\perp})+{\not\!\epsilon}_{\perp}{\not\!P}\psi_6(q_{\perp})$. If we further set $\psi_5(q_{\perp})=-\psi_6(q_{\perp})=\psi(q_{\perp})$, the wave function reduces to the non-relativistic case, e.g. Eq. (\ref{eq:schrodinger}). Thus, the terms with $\psi_1$, $\psi_2$, $\psi_3$, $\psi_4$, $\psi_7$ and $\psi_8$  in Eq. (\ref{eq3-2}) are all relativistic corrections.

When the charge conjugate parity is taken into account, the terms with $\psi_2(q_{\perp})$ and $\psi_7(q_{\perp})$ vanish because of the positive charge conjugate parity $c=+$, and the general instantaneous wave function for the $1^{--}$ quarkonium becomes
$$\Psi_{_P}^{1^{--}}(q_{\perp})=
q_{\perp}\cdot{\epsilon}_{\perp}
\left[\psi_1(q_{\perp})+
\frac{{\not\!q}_{\perp}}{M}\psi_3(q_{\perp})+\frac{{\not\!P}
{\not\!q}_{\perp}}{M^2} \psi_4(q_{\perp})\right]+
M{\not\!\epsilon}_{\perp}\psi_5(q_{\perp})$$
\begin{equation}+
{\not\!\epsilon}_{\perp}{\not\!P}\psi_6(q_{\perp})+\frac{1}{M}({\not\!P}{\not\!\epsilon}_{\perp}
{\not\!q}_{\perp}-{\not\!P}q_{\perp}\cdot{\epsilon}_{\perp})
\psi_8(q_{\perp}).\label{eq3-3}
\end{equation}

Before moving on, we would like to discuss the interaction kernel ${{V(r)}}$. We know from Quantum Chromodynamics that the strong interaction between a quark and antiquark is given by the exchange of gluon(s), and that the basic kernel contains a short-range $\gamma_{\mu}\otimes\gamma^{\mu}$ vector interaction $-\frac{4\alpha_s}{3r}$ plus a long-range $1\otimes 1$ linear confining scalar interaction $\lambda r$ suggested by the lattice QCD calculations \cite{godfrey}. In the Coulomb gauge and in the leading order, the kernel is the famous Cornell potential
\begin{equation}
V(r)=\lambda r+V_0-\gamma_0\otimes\gamma^0\frac{4}{3}\frac{\alpha_s}{r},\label{potential}
\end{equation}
where $\lambda$ is the string tension, $V_0$ is a free constant appearing in the potential to fit the data, and $\alpha_s$ is the
running coupling constant. In order to avoid infrared divergence and incorporate the screening effects, an exponential factor $e^{-\alpha r}$ is added to the potential \cite{laermann}, i.e.
\begin{equation}
V(r)= \frac{\lambda}{\alpha}(1-e^{-\alpha r})+V_0-\gamma_0\otimes\gamma^0\frac{4}{3}\frac{\alpha_s}{r}e^{-\alpha r}.
\label{potential1}
\end{equation}
It is easy to check that when $\alpha r\ll 1$, Eq. (\ref{potential1}) reduces to Eq. (\ref{potential}). In the momentum space and in the rest frame of the bound state, the potential takes the form:
\begin{equation}\label{potential2}
V(\vec{q})=V_s(\vec{q})+\gamma_{_0}\otimes\gamma^0 V_v(\vec{q}),
\end{equation}
where$$V_s(\vec{q})=-(\frac{\lambda}{\alpha}+V_0)
\delta^3(\vec{q})+\frac{\lambda}{\pi^2}
\frac{1}{(\vec{q}^2+{\alpha}^2)^2},~~
V_v(\vec{q})=-\frac{2}{3{\pi}^2}\frac{\alpha_s(
\vec{q})}{{(\vec{q}}^2+{\alpha}^2)},$$
$$\alpha_s(\vec{q})=\frac{12\pi}{33-2N_f}\frac{1}
{\log (e+\frac{{\vec{q}}^2}{\Lambda^{2}_{QCD}})}.$$
Here, $\alpha_s(\vec{q})$ is the running coupling of the one loop QCD correction, and $e=2.71828$. The constants $\lambda$, $\alpha$, $V_0$ and $\Lambda_{QCD}$ are the parameters which characterize the potential, and $N_f=3$ for the $c \bar c$ system, $N_f=4$ for the $b \bar b$ system.

The reader may wonder why we have chosen a simple basic kernel, and not a relativistic one \cite{godfrey, ebert} which includes details of the spin-independent potential and the spin-dependent potential, like the spin-spin interaction, spin-orbital interaction, tensor interaction, etc. The reason is that in our relativistic method, with a relativistic wave function for the bound state, we only need the basic potential and not a relativistic one, otherwise we would have double counting. To explain this, let us show how the relativistic potential is obtained: the potential between a quark and anti-quark is constructed from the on-shell $q \bar q$ scattering amplitude in the center-of-mass frame motivated by single gluon exchange, where the gluon propagator is given in the Coulomb gauge. The basic non-relativistic vector potential $-\frac{4\alpha_s}{3r}$ is obtained at leading-order from the amplitude (usually in the momentum space). To obtain the relativistic corrections of the potential, the on-shell Dirac spinors of the quark and anti-quark are expanded in quantities like the mass, momentum, etc. The relativistic potential is then obtained, and the relativistic corrections from the free spinors (wave functions for a bound state) are moved to the potential. The corresponding wave function becomes non-relativistic.

In our case, we have a relativistic wave function and the potential is non-relativistic. If both of them are relativistic, then there is double counting. In general, a relativistic method should have a relativistic wave function with a non-relativistic potential, or a non-relativistic wave function with a relativistic potential. In principle, a half-relativistic wave function with a half-relativistic potential is also permitted, but one has to be careful to avoid double counting. The method with a non-relativistic wave function and a relativistic potential is usually good for calculating the mass spectrum of bound state, while the method with a relativistic wave function and a non-relativistic potential is not only good for calculating the mass spectrum as an eigenvalue problem, but is also good for calculating the transition amplitude.

With the kernel Eq. (\ref{potential2}) and the relativistic wave function Eq. (\ref{eq3-2}) or Eq. (\ref{eq3-3}), we are ready to solve the coupled Salpeter equation Eqs. (11-14). Substituting the wave function Eq. (\ref{eq3-3}) into Eq. (13) and Eq. (14), taking the trace on both sides, multiplying with the polarization vector on both sides, e.g. $q_{\perp}\cdot \epsilon^{*}$ or ${\slashed{\epsilon}^{*}_{\perp}}\cdot \slashed{P}$, and then using the completeness of the polarization vector, we obtain the relations $$\psi_1(q_{\perp})=\frac{q_{\perp}^2 \psi_3(q_{\perp}) +M^2\psi_5(q_{\perp})}
{Mm_1},\,~~~\psi_8(q_{\perp})=-\frac{\psi_6(q_{\perp})M} {m_1},$$
where we have used $m_1=m_2$ for a quarkonium state. We now have only four independent unknown radial wave functions, $\psi_3(q_{\perp})$, $\psi_4(q_{\perp})$, $\psi_5(q_{\perp})$, $\psi_6(q_{\perp})$, whose numerical values can be obtained by solving Eq. (11) and Eq. (12). Substituting the wave function Eq. (\ref{eq3-3}) into Eq. (11) and Eq. (12), and taking the trace again, we finally obtain four coupled equations
\begin{eqnarray}
&\displaystyle
(M-2\omega_1)\left\{\left(\psi_3(\vec {q})\frac{\vec {q}^2}{M^2}-\psi_5(\vec {q})\right)
+\left(\psi_4(\vec {q})\frac{\vec {q}^2}{M^2}+\psi_6(\vec {q})\right)\frac{m_1}
{\omega_1}\right\}\nonumber\\
&\displaystyle=\int{\frac{d^3\vec{k}}{(2\pi)^3}\frac{2}{\omega_1^2}}\left\{(V_s+V_v)
\left(\psi_3(\vec{k})\frac{{\vec{k}}^2}{M^2}-\psi_5(\vec{k})\right)
(\vec{k}\cdot\vec{q})\right.\nonumber\\
&\displaystyle-(V_s-V_v)\left[m_1^2\left(\psi_3(\vec{k})\frac{(\vec{k}\cdot
\vec{q})^2}{M^2\vec {q}^2}-\psi_5(\vec{k})\right)\left.+m_1\omega_1\left(\psi_4(\vec{k})\frac{(\vec{k}\cdot
\vec{q})^2}{M^2\vec {q}^2}+\psi_6(\vec{k})\right)\right]\right\}\,,
\end{eqnarray}
$$(M+2\omega_1)\left\{\left(\psi_3(\vec {q})\frac{\vec {q}^2}{M^2}-\psi_5(\vec {q})\right)
-\left(\psi_4(\vec {q})\frac{\vec {q}^2}{M^2}+\psi_6(\vec {q})\right)\frac{m_1}
{\omega_1}\right\}$$
$$=-\int{\frac{d^3\vec{k}}{(2\pi)^3}\frac{2}{\omega_1^2}}\left\{(V_s+V_v)\left[
\left(\psi_3(\vec{k})\frac{{\vec{k}}^2}{M^2}-\psi_5(\vec{k})\right)
\right](\vec{k}\cdot\vec{q})\right.$$
\begin{equation}-(V_s-V_v)\left[m_1^2\left(\psi_3(\vec{k})\frac{(\vec{k}\cdot
\vec{q})^2}{M^2\vec {q}^2}-\psi_5(\vec{k})\right)\left.-m_1\omega_1\left(\psi_4(\vec{k})\frac{(\vec{k}\cdot
\vec{q})^2}{M^2\vec {q}^2}+\psi_6(\vec{k})\right)\right]\right\}\,,
\end{equation}
$$(M-2\omega_1)\left\{\left(\psi_3(\vec {q})+\psi_4(\vec {q})\frac{m_1}{\omega_1}\right)
\frac{\vec {q}^2}{M^2} -3\left(\psi_5(\vec {q})-\psi_6(\vec {q})\frac
{\omega_1}{m_1}\right)-\psi_6(\vec {q})\frac{\vec {q}^2}{m_1\omega_1}\right\}$$
$$
=-\int{\frac{d^3\vec{k}}{(2\pi)^3}\frac{1}{\omega_1^2}}\left\{(V_s+V_v)
\left[-\frac{2\omega_1}{m_1} \psi_6(\vec{k})-\psi_3(\vec{k})\frac{{\vec k}^2}{M^2}+
\psi_5(\vec{k})\right](\vec{k}\cdot \vec{q})\right.$$
$$
+(V_s-V_v)\left[\omega_1^2\left(\psi_3(\vec{k})\frac{{\vec k}^2}{M^2}
-3\psi_5(\vec{k})\right)+m_1\omega_1
\left(\psi_4(\vec{k})\frac{{\vec k}^2}{M^2}+3\psi_6(\vec{k})\right)\right.$$
\begin{equation}
\left.\left. -\left(\psi_3(\vec{k})\frac{(\vec{k}\cdot\vec{q})^2}{M^2}
-\psi_5(\vec{k})\vec{q}^2\right)\right]\right\}\,,
\end{equation}
$$(M+2\omega_1)\left\{\left[\psi_3(\vec {q})-\psi_4(\vec {q})\frac{m_1}{\omega_1}\right]
\frac{\vec {q}^2}{M^2}-3\left(\psi_5(\vec {q})+\psi_6(\vec {q})\frac
{\omega_1}{m_1}\right)+\psi_6(\vec {q})\frac{\vec {q}^2}{m_1\omega_1}\right\}$$
$$
=\int{\frac{d^3\vec{k}}{(2\pi)^3}\frac{1}{\omega_1^2}}\left\{(V_s+V_v)
\left[\frac{2\omega_1}{m_1}\psi_6(\vec{k}) -\psi_3(\vec{k})\frac{{\vec k}^2}{M^2}+
\psi_5(\vec{k})\right](\vec{k}\cdot\vec{q})\right.$$
$$
+(V_s-V_v)\left[\omega_1^2\left(\psi_3(\vec{k})\frac{{\vec k}^2}{M^2}-3\psi_5(\vec{k})\right)-m_1\omega_1
\left(\psi_4(\vec{k})\frac{{\vec k}^2}{M^2}+3\psi_6(\vec{k})\right)\right.$$
\begin{equation}
\left.\left. -\left(\psi_3(\vec{k})\frac{(\vec{k}\cdot\vec{q})^2}{M^2}
-\psi_5(\vec{k}){\vec {q}^2}\right)\right]\right\}\,,
\end{equation}
where we have used the relation $\omega_1=\omega_2$ for a quarkonium, and $V_s=V_s(\vec q - \vec k)$, $V_v=V_v(\vec q - \vec k)$. Since we have four coupled equations, the four independent radial wave functions can be obtained numerically, and the mass spectrum obtained simultaneously as the as the eigenvalue problem.

The normalization condition Eq. (\ref{nor}) for the $1^{--}$ wave function is
$$
\int \frac{d^3{\vec q}}{(2\pi)^3}\frac{8\omega_1}{3M}\left\{
3\psi_5(\vec {q})\psi_6(\vec {q})\frac{M^2}{2m_1}\right.
$$
\begin{equation}\left.+\frac{\vec q^2}{2m_1}\left[
\psi_4(\vec {q})\psi_5(\vec {q})-\psi_3(\vec {q})\left(\psi_4(\vec
{q})\frac{\vec q^2}{M^2}+\psi_6(\vec {q})\right)\right]
\right\}=1.
\end{equation}

\section{The Decay Constant in the Salpeter Method}

The relativistic decay constant $F^{Re}_{V}$ in Eq. (\ref{decayconstant}) for a vector quarkonium can be calculated in the BS method as
\begin{equation}
F^{Re}_V M\epsilon_{\mu}=\sqrt{N_c}\int\frac{d^4q}{(2\pi)^4}\mathrm{Tr}[\chi_{_P}({q})\gamma_\mu]
=i\sqrt{N_c}\int\frac{d^3{\vec q}}{(2\pi)^3}\mathrm{Tr}[\Psi_{_P}({\vec q})\gamma_\mu],
\end{equation}
where $N_c=3$ is the color number, and ${{\mathrm{Tr}}}$ is the trace operator. We note that when calculating the decay constant, the Salpeter wave function $\Psi_{_P}({\vec q})$, and not merely the positive wave function $\Psi^{++}_{_P}({\vec q})$ gives a contribution. For a vector quarkonium with the relativistic wave function Eq. (\ref{eq3-3}), we obtain the relativistic decay constant
\begin{eqnarray}
F^{Re}_{V} = 4\sqrt{3} \int \frac{d^3 \vec{q}}{(2\pi)^3} \left[\psi_{5}({\vec
q})-\frac{{\vec q}^2}{3M^2}\psi_3(\vec{q})\right],\label{vectorFV}
\end{eqnarray}
where we note that the $\psi_{5}$ and $\psi_{3}$ term both contribute.

\section{Results and Discussion}

\subsection{Input parameters and the heavy quarkonium wave functions}

The input parameters can be fixed by fitting the mass spectra of charmonium and bottomonium. We choose $m_b=4.96~\mathrm{GeV}$, $m_c=1.60~\mathrm{GeV}$, $\alpha=0.06$ GeV, and $\Lambda_{\rm QCD}=0.21$ GeV \footnote{In previous Letter \cite{fu}, we have chosen different $\Lambda_{\rm QCD}$ for charmonium and bottomonium. Since this parameter appears only in $\alpha_s$, which depends on $\vec q$, and for more convenience of fitting the data, we choose the same $\Lambda_{\rm QCD}$ for the two systems.}. We also choose $\lambda=0.23$ GeV$^2$ and $V_0 =-0.249$ GeV for the charmonium system, and $\lambda=0.2$ GeV$^2$ and $V_0 =-0.124$ GeV for the bottomonium system.

\begin{table}[htb]
\begin{center}
\caption{Mass spectra of the $S$ wave $c\bar c$ and $b\bar b$
vectors in units of MeV. `Th' is the theoretical prediction, `Exp' are the experimental data from PDG
\cite{pdg}.} \vspace{0.5cm}\label{table1}
\begin{tabular}
{|c|c|c|c|c|c|c|}\hline ~~~$ nS$~~~&~~~Th($c\bar
c$)~~~&~~~Exp($c\bar c$)~~~&~~~Th($b\bar b$)~~~&~~~Exp($b\bar b$)~~~
\\\hline
$~1S$&3097.3&3096.9&9460.7&9460.3
\\\hline
$~2S$&3686.4&3686.1&10020.5&10023.3
\\\hline
$~3S$&4059.3&4039&10362.6&10355.2
\\\hline
$~4S$&4337.5&4421&10622.2&10579.4
\\\hline
$~5S$&4559.4& / &10835.1&10889.9 \\\hline
\end{tabular}
\end{center}
\end{table}

\begin{figure}[htb]
\begin{picture}(260,240)(60,30)
\put(0,0){\includegraphics{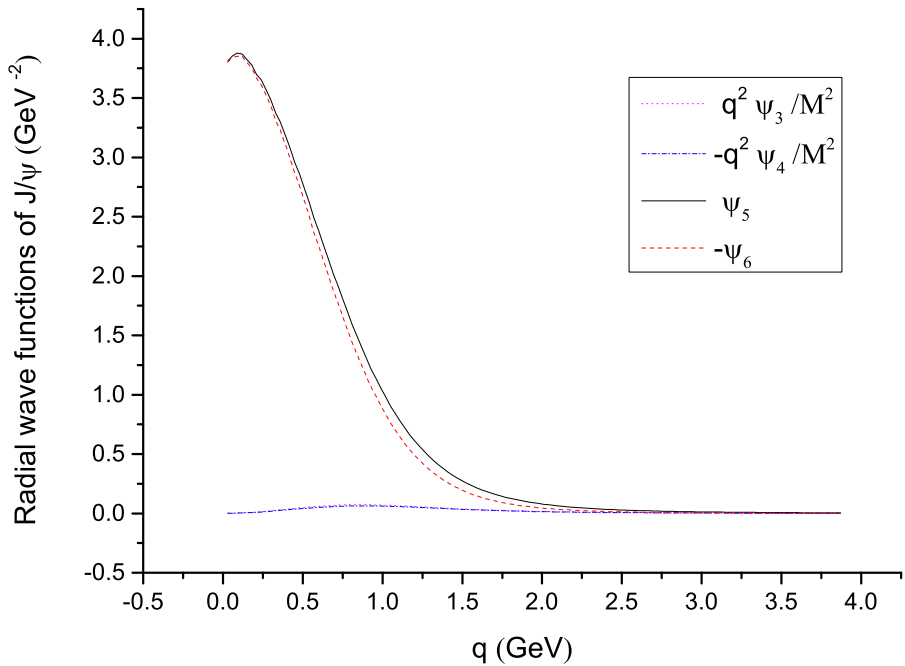}}
\end{picture}
\caption{Four typical radial wave functions of $J/\psi$.
}\label{figure1}
\end{figure}

The mass spectra of vector charmonium and bottomonium are shown in Table \ref{table1}. The theoretical predictions are consistent with the experimental data given by the Particle Data Group (PDG). An an example of the wave functions, we present four $J/\psi$ radial wave functions in Figure \ref{figure1}: the dominant radial wave functions $\psi_5$ and $\psi_6$ and the two minor ones ${\vec q}^2\psi_3/M^2$ and ${\vec q}^2\psi_4/M^2$ \footnote{Here we show the curves of ${\vec q}^2\psi_3/M^2$ and ${\vec q}^2\psi_4/M^2$ other than $\psi_3$ and $\psi_4$, because they always appear in such a combined form in the applications.}. From now on, we use the symbols $|\vec q|=q$ and $|\vec v|=v$ for simplicity.

\begin{figure}[htb]
\begin{picture}(260,240)(60,30)
\put(0,0){\includegraphics{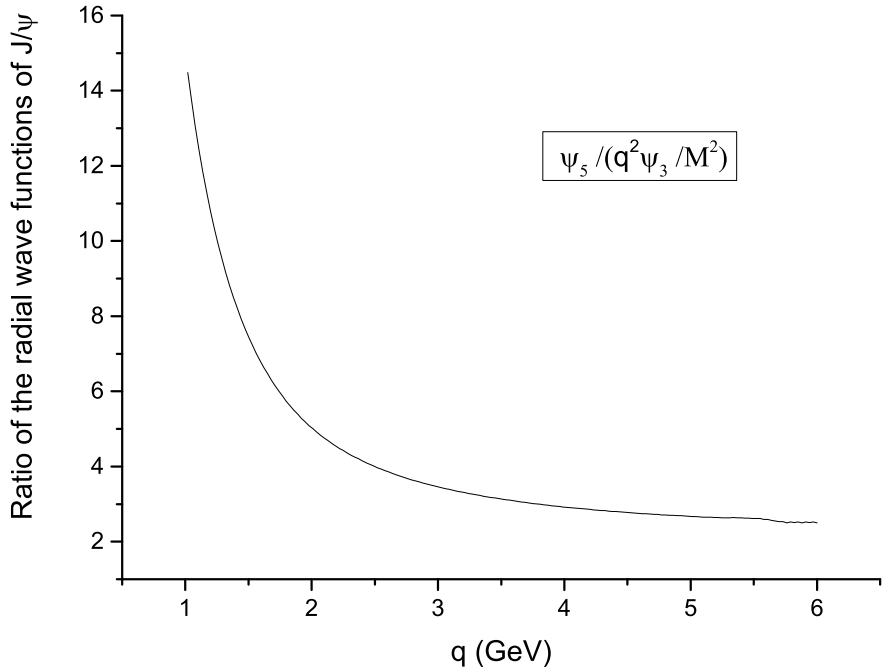}}
\end{picture}
\caption{Ratio $\frac{\psi_5}{{q}^2\psi_3/M^2}$ of the
$J/\psi$ radial wave functions. }\label{figure2}
\end{figure}

As described in Sec 4, the terms with radial wave functions $\psi_5$ and $\psi_6$ in the total wave function Eq.(\ref{eq3-3}), are non-relativistic, while all the others are relativistic corrections. Figure \ref{figure1} shows that the relativistic wave functions $\psi_3$ and $\psi_4$ are small and could be safely neglected, but in fact this is the case. Figure \ref{figure1} only shows the relative importance of the wave functions in the region of small, and to see the relative importance of the wave functions in the whole $q$ region, we plot the ratio $\psi_5/({q}^2\psi_3/M^2)$ in Figure \ref{figure2}. It can be seen that in the large $q$ region, the value of $\psi_5$ is only a few times larger than of $({q}^2\psi_3/M^2)$. Thus, the terms which are proportional to $\psi_{3}$ ($\psi_{4}$ and others) may have a sizable contribution in the large $q$ region, leading to possibly important relativistic corrections.

\subsection{Charmonium leptonic decay widths}

\begin{table}[htb]
\begin{center}
\caption{Decay rates of $\psi(nS)\to\ell^+\ell^-$ in units of
keV. `NR' is the non-relativistic result, `Re' is the relativistic result,
`Exp' are the experimental data from PDG \cite{pdg}.}
\vspace{0.5cm}\label{table2}
\begin{tabular}
{|c|c|c|c|c|}\hline ~~~~modes~~~~&~~~~NR~~~~&~~~~
Re~~~~&~~$\frac{\rm{NR-Re}}{\rm{Re}}$~~&~~~~Exp~~~~
\\\hline
$J/\psi\to
e^{+}e^{-}$&$10.95^{+2.20}_{-1.86}$&$8.95^{+1.57}_{-1.38}$&$22.3^{+2.7}_{-2.2}\%$&~~5.55$\pm$0.16~~
\\\hline
$\psi(2S)\to e^{+}e^{-}$&$5.92^{+1.05}_{-0.89}$&$4.43^{+0.60}_{-0.54}$&$33.6^{+5.0}_{-4.3}\%$&$2.33\pm$0.04
\\
$\psi(2S)\to
\tau^{+}\tau^{-}$&$2.31^{+1.68}_{-2.31}$&$1.73^{+1.29}_{-1.73}$&$33.6^{+3.2}_{-4.1}\%$&0.91$\pm$0.14
\\\hline
$\psi(3S)\to e^{+}e^{-}$&$4.30^{+0.69}_{-0.66}$&$3.04^{+0.35}_{-0.35}$&$41.4^{+5.6}_{-6.1}\%$&0.86$\pm$0.07
\\
$\psi(3S)\to\tau^{+}\tau^{-}$&$2.87^{+0.61}_{-1.23}$&$2.03^{+0.48}_{-0.87}$&$41.4^{+6.9}_{-6.3}\%$&   /
\\\hline

$\psi(4S)\to e^{+}e^{-}$&$3.53^{+0.65}_{-0.66}$&$2.32^{+0.24}_{-0.26}$&$52.2^{+11.1}_{-12.9}\%$&0.48$\pm$0.22
\\
$\psi(4S)\to\tau^{+}\tau^{-}$&$2.70^{+0.26}_{-0.60}$&$1.78^{+0.25}_{-0.43}$&$52.2^{+15.4}_{-13.0}\%$&   /
\\\hline
$\psi(5S)\to e^{+}e^{-}$&$3.05^{+0.55}_{-0.52}$&$1.88^{+0.16}_{-0.19}$&$62.2^{+14.3}_{-12.5}\%$&0.58$\pm$0.07
\\
$\psi(5S)\to\tau^{+}\tau^{-}$&$2.49^{+0.30}_{-0.58}$&$1.54^{+0.21}_{-0.31}$&$62.2^{+16.4}_{-23.1}\%$&   /
\\\hline
\end{tabular}
\end{center}
\end{table}

Our results for $\psi(nS)\to\ell^+\ell^-$ are shown in Table \ref{table2}, where in the second column, `NR', the non-relativistic decay rates are shown, meaning that in Eq. (\ref{vectorFV}) the $\psi_3$ term is ignored, so that the only contribution is from the $\psi_5$ tern. The third column, `Re', show the relativistic results
including the contributions of $\psi_5$ and $\psi_3$. One can see that for charmonium the relativistic results are different from the non-relativistic ones. To see this clearly, we add the fourth column in Table \ref{table2} with the ratio (NR-Re)/Re, whose value can be called the  `relativistic effect'.

\begin{figure}
\begin{picture}(260,240)(60,30)
\put(0,0){\includegraphics{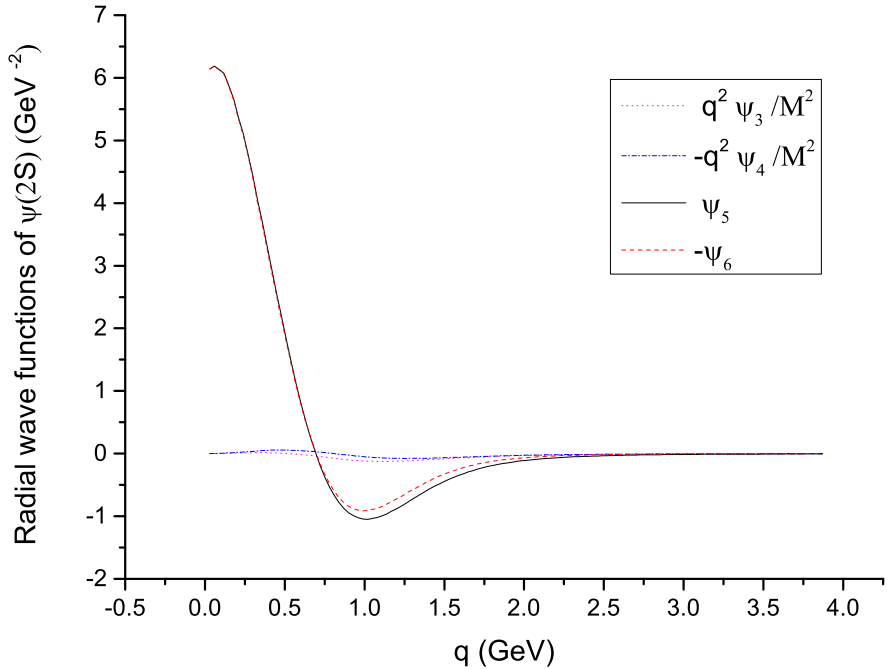}}
\end{picture}
\caption{Radial wave functions of $\psi(2S)$.
}\label{figure3}
\end{figure}

Table \ref{table2} indicates that the relativistic effect is about $22\%$ for the $J/\psi$ decay, which is consistent with the usual power relation for the relativistic terms, e.g. $v_c^2 \sim 0.2-0.3$. For the excited states, the relativistic effects are much larger than for the ground state. For the $2S$, $3S$, $4S$ and $5S$ states, the relativistic effects are about $34\%$, $41\%$, $52\%$ and $62\%$, respectively. These results are consistent with our previous study of the semi-leptonic decays $B^+_c \to {c\bar c}+\ell^+ +\nu_{\ell}$, where higher excited charmonium states were shown to have larger relativistic effects \cite{geng}. This conclusion can also be obtained qualitatively from the plots of radial wave functions. We mentioned that the relative momentum $q$ concerns the relative velocity $v_Q$ between the quark and antiquark in quarkonium, $q =0.5 m_Q  v_Q$. As shown in Figure \ref{figure1}, two non-relativistic $J/\psi$ radial wave functions always dominate over the relativistic wave functions in the whole $q$ region, leading to a small relativistic correction. For the excited states, see Figure \ref{figure3} as an example of the radial wave functions of $\psi(2S)$, the non-relativistic wave functions still dominate in the small $q$ region, but there is a node structure in each curve where the wave function changes sign. The contributions in the low $q$ region may cancel each other, and the wave functions for large $q$ ($v_Q$) may give sizable contributions, resulting in large relativistic correction.

There are other methods for considering the relativistic effects in heavy quarkonium decays. For example, Bodwin {\it et al.}~\cite{bodwin} and Brambilla {\it et al.}~\cite{brambilla} computed the ${v_Q}^2$ and the ${v_Q}^4$ corrections of the decay rate of $Q\bar{Q}$ quarkonium in the framework of NRQCD. In the case of $J/\psi$~\cite{bodwin}, the predicted relativistic effect is $34.1\%$ for $v_c^2\sim 0.3$, and is $23.0\%$ for ${v_c}^2\sim 0.18$. These values are consistent with our prediction of $22.3\%$.

In Table \ref{table2}, we also show the theoretical uncertainties caused by the choice of input parameters. We vary all parameters simultaneously within $\pm 10\%$ of their central values, and take the largest variation as the uncertainty. With the errors, most predictions are much larger than the experimental data. The only exception is the channel $\psi(2S)\to \tau^+\tau^{-}$, which has large uncertainties \footnote{The reason is that the $\psi(2S)$ mass is only a slightly heavier than that of two $\tau$, so the phase space of this channel is very sensitive to the variation of parameters.}. We note that a calculation of the $J/\psi$ leptonic decay in lattice QCD with fully relativistic charm quarks was reported in Ref. \cite{donald}, and give $\Gamma(J/\psi\to e^+e^-)=5.48(16)$ keV, consistent with the experimental data. This indicates that the disagreement of our results with the experimental data may be due to the lack of QCD corrections.

We note that in a recent paper, Soni {\it et al.}~\cite{soni} calculated the quarkonium leptonic decay using the Cornell potential in a non-relativistic version and with the pQCD correction up to NLO. Their results for charmonium are neither consistent with the experimental data nor with our results, while for bottomonium their results are comparable with ours (see below). Also, Badalian {\it et al.}~\cite{badalian1} calculated the decay rates of the $\psi(1S-4S)$ leptonic decays with QCD correction at NLO using the Cornell potential and the semi-Salpeter equation, and obtained 5.47 keV, 2.68 keV, 1.97 keV, and $1.58$ keV, respectively, which are smaller than our charmonium results. These studies indicate that the relativistic corrections and QCD corrections are large for the charmonium system.

\subsection{Bottomonium leptonic decay widths}

\begin{table}[htb]
\begin{center}
\caption{Decay rates of $\Upsilon(nS)\to\ell^+\ell^-$ in units
of keV. `NR' is the non-relativistic result, `Re' is the relativistic result,
`Exp' are the experimental data \cite{pdg}.}
\vspace{0.5cm}\label{table3}
\begin{tabular}
{|c|c|c|c|c|c|c|}\hline~~~~ ~~~~modes~~~~~~~~&~~~~~~~NR~~~~~~~&~~~~~~~
Re~~~~~~~&~~~~$\frac{\rm{NR-Re}}{\rm{Re}}$~~~~&~~~~Exp~~~~
\\\hline
$\Upsilon(1S)\to
e^+e^-$&1.47$^{+0.23}_{-0.20}$&1.29$^{+0.19}_{-0.16}$&14.0$^{+0.9}_{-1.6}$\%&
1.340$\pm$0.018 (1.29$\pm$0.09)
\\
$\Upsilon(1S)\to\tau^{+}\tau^{-}$&1.46$^{+0.22}_{-0.20}$&1.28$^{+0.18}_{-0.16}$&14.0$^{+1.1}_{-1.5}$\%&
1.40$\pm$0.09
\\\hline
$\Upsilon(2S)\to
e^+e^-$&0.771$^{+0.123}_{-0.125}$&0.629$^{+0.104}_{-0.088}$&22.6$^{+0.0}_{-3.2}$\%&
0.612$\pm$0.011
\\
$\Upsilon(2S)\to\tau^{+}\tau^{-}$&0.766$^{+0.120}_{-0.123}$&0.625$^{+0.101}_{-0.086}$&22.6$^{+0.0}_{-3.3}$\%&
0.64$\pm$0.12
\\\hline
$\Upsilon(3S)\to
e^+e^-$&0.541$^{+0.088}_{-0.088}$&0.450$^{+0.070}_{-0.065}$&20.2$^{+7.8}_{-2.5}$\%&
0.443$\pm$0.008
\\
$\Upsilon(3S)\to\tau^{+}\tau^{-}$&0.538$^{+0.086}_{-0.087}$&0.448$^{+0.068}_{-0.064}$&20.2$^{+7.7}_{-2.8}$\%&
0.47$\pm$0.10
\\\hline
$\Upsilon(4S)\to
e^+e^-$&0.429$^{+0.083}_{-0.059}$&0.355$^{+0.058}_{-0.050}$&20.8$^{+6.6}_{-7.2}$\%&
~~0.272$\pm$0.029~(0.322$\pm$0.056)~~
\\
$\Upsilon(4S)\to\tau^{+}\tau^{-}$&0.427$^{+0.081}_{-0.057}$&0.353$^{+0.056}_{-0.050}$&20.8$^{+6.5}_{-7.1}$\%&  /

\\\hline
$\Upsilon(5S)\to
e^+e^-$&0.380$^{+0.048}_{-0.069}$&0.296$^{+0.048}_{-0.038}$&28.4$^{+1.1}_{-7.9}$\%&
0.31$\pm$0.07
\\
$\Upsilon(5S)\to\tau^{+}\tau^{-}$&0.378$^{+0.047}_{-0.068}$&0.295$^{+0.047}_{-0.038}$&28.4$^{+1.0}_{-7.8}$\%&  /

\\\hline
\end{tabular}
\end{center}
\end{table}

We present the non-relativistic and relativistic results of the bottomonium leptonic decay widths in Table \ref{table3}. Similarly to charmonium, the relativistic corrections are also sizable. For the ground state $\Upsilon(1S)$, the relativistic effect is about $14\%$, and for the excited states $\Upsilon(2S-5S)$, they vary from $20\%$ to $28\%$. These predictions agree with those in literature. For example, Bodwin {\it et al.} \cite{bodwin} predicted the relativistic effect of $13.2\%$ for ${v_b}^2\sim0.10$ using NRQCD up to the ${v_b}^4$ accuracy, and a lattice QCD prediction indicated that the relativistic effects are about $(15-25)\%$~\cite{colquhoun} for $\Upsilon(1S)$ and $\Upsilon(2S)$ up to the ${v_b}^2$ accuracy.

We should point out that the above large relativistic effects are specific for bottomonium leptonic decays $\Upsilon(nS) \to \ell^+\ell^-$, and are not universal for processes involving a bottomonium. In the di-lepton decays, the amplitude is proportional to the wave function as $\int {d^3 \vec{q}} \left[\psi_{5}({\vec q})-\frac{{\vec q}^2}{3M^2}\psi_3(\vec{q})\right]$, i.e. the wave function is to the power of one. For other processes, such as meson $A$ to meson $B$ semileptonic decays, the amplitude is proportional to the overlapping integral of the wave functions for the initial and final states $\int {d^3 \vec{q}}~ \psi_A \cdot \psi_B$. Because the wave functions are large in the small $q$ region, the product of two wave functions is suppressed in the large $q$ region compared to the case with one wave function, and the contributions from the relativistic terms are greatly suppressed.

In Table \ref{table3}, we also give the theoretical uncertainties, which are obtained by varying all parameters simultaneously within $\pm 10\%$ of the central values, and the largest variations are taken as the errors. Our relativistic results agree well with the experimental data. We also note that, for $\Upsilon(1S) \to e^+e^-$, PDG gives two different results: directly listed is $\Gamma_{ee}=1.34$ keV, but a branching ratio $Br=2.38\%$ is also given, leading to $\Gamma_{\Upsilon(1S)\to e^+e^-}=1.29$ keV using the full width $\Gamma_{\Upsilon(1S)}=54.02$ keV \cite{pdg}. The second value is the same as our relativistic result. Similarly, in the case of $\Upsilon(4S) \to e^+e^-$, PDG directly lists $\Gamma_{ee}=0.272$ keV \cite{pdg}, but from the branching ratio also given in PDG, we get $\Gamma_{ee}=0.322$ keV. We hope PDG will update the data in the near future.

Table \ref{table3} shows that all relativistic results for $\Gamma_{\ell\ell}(1S-5S)$ are consistent with the experimental data. Our predictions also agree with the lattice QCD prediction~\cite{colquhoun}, $\Gamma(\Upsilon(1S)\to e^+e^-)=1.19(11)$ keV and $\Gamma(\Upsilon(2S)\to e^+e^-)=0.69(9)$ keV, with the NRQCD prediction~\cite {pineda2}, $\Gamma(\Upsilon(1S)\to e^+e^-)=1.25$ keV, and with the NRQCD prediction with NNNLO pQCD corrections~\cite{beneke2}, $\Gamma(\Upsilon(1S)\to e^+e^-)=1.08\pm0.05(\alpha_s)^{+0.01}_{-0.20}(\mu)$ keV.

\subsection{Lepton flavor university}

\begin{table}[htb]
\begin{center}
\caption{Ratios 
$R^{\psi_{nS}}_{\tau\tau}=\frac{\Gamma(\psi_{nS}\to\tau^{+}\tau^{-})}{\Gamma(\psi_{nS}\to
\mu^{+}\mu^{-})}$ and $R^{\Upsilon_{nS}}_{\tau\tau}$. The
experimental data are from PDG \cite{pdg}, with the
statistical and systematic uncertainties added together.}
\vspace{0.5cm}\label{table4}
\begin{tabular}
{|c|c|c|c|c|c|}\hline
~~~~~~&~~~~~~&~~~$R^{\psi_{2S}}_{\tau\tau}$~~~&~~~$R^{\psi_{3S}}_{\tau\tau}$~~~&~~~
$R^{\psi_{4S}}_{\tau\tau}$~~~&~~~$R^{\psi_{5S}}_{\tau\tau}$~~~
\\\hline
Ours&&0.391$^{+0.210}_{-0.391}$&0.668$^{+0.073}_{-0.237}$&0.767$^{+0.026}_{-0.112}$&0.819$^{+0.039}_{-0.091}$
\\\hline
~~~~~~&~~~$R^{\Upsilon_{1S}}_{\tau\tau}$~~~&~~~$R^{\Upsilon_{2S}}_{\tau\tau}$~~~&~~~$R^{\Upsilon_{3S}}_{\tau\tau}$~~~&~~~
$R^{\Upsilon_{4S}}_{\tau\tau}$~~~&~~~$R^{\Upsilon_{5S}}_{\tau\tau}$~~~
\\\hline
Ours&0.992$^{+0.001}_{-0.006}$&0.994$^{+0.003}_{-0.003}$&0.996$^{+0.002}_{-0.002}$&0.995$^{+0.001}_{-0.002}$&0.997$^{+0.000}_{-0.002}$
\\
~~CLEO\cite{besson}~~&~~$1.02\pm0.07$~~&~~$1.04\pm0.09$~~&~~$
1.05\pm0.13$~~&  /&  /
\\
~~BABAR\cite{sanchez}~~&~~$1.005\pm0.035$~~&  / &  / &   /   &   /
\\
~~PDG~\cite{pdg}~~&~~$1.05\pm0.06$~~&$1.04\pm0.20$&$1.05\pm0.24$&  / &   /
\\\hline
\end{tabular}
\end{center}
\end{table}

To test the lepton flavor university, we give the ratios $R^{\psi_{nS}}_{\tau\tau}$ and $R^{\Upsilon_{nS}}_{\tau\tau}$ in Table \ref{table4}. Their definitions are similar, for example, $$R^{\psi_{nS}}_{\tau\tau}=\frac{\Gamma(\psi_{nS}\to\tau^{+}\tau^{-})}{\Gamma(\psi_{nS}\to
\mu^{+}\mu^{-})}.$$ The deviation of the ratio $R^{\psi_{nS}}_{\tau\tau}$ from the lepton flavor universality indicates the presence of new physics beyond the Standard Model.

Table \ref{table4} shows the ratios calculated with the `Re' values. The uncertainties of the ratio are from the variation of the input parameters. In the case of charmonium, the ratios $R^{\psi_{nS}}_{\tau\tau}$ are quite different from each other, since the charmonium mass is a bit higher than of two $\tau$. For the same reason, we get a large uncertainty. For bottomonium since the $\tau$ mass is much smaller than the bottomonium mass, we get almost the same values for all ratios $R^{\Upsilon_{nS}}_{\tau\tau}$. Their uncertainty is also very small due to the cancellation between the numerator and denominator. Even though all central values of the ratios $R^{\Upsilon_{nS}}_{\tau\tau}$ are smaller than $1$, they are consistent with the existing experimental data within errors.

\begin{table}[htb]
\begin{center}
\caption{Ratio $\Gamma(\psi(nS)\to e^+e^-)/\Gamma(J/\psi\to e^+e^-)$.} \vspace{0.5cm}
\label{table5}
\begin{tabular}
{|c|c|c|c|c|}\hline
~~~~~~~~~~~~&~~~$\frac{\Gamma(\psi(2S))}{\Gamma(J/\psi)}$~~~
&~~~$\frac{\Gamma(\psi(3S))}{\Gamma(J/\psi)}$~~~
&~~~$\frac{\Gamma(\psi(4S))}{\Gamma(J/\psi)}$~~~
&~~~$\frac{\Gamma(\psi(5S))}{\Gamma(J/\psi)}$~~~
\\\hline
Ours&0.495$^{+0.019}_{-0.017}$&0.340$^{+0.016}_{-0.017}$&0.259$^{+0.013}_{-0.016}$&0.210$^{+0.013}_{-0.016}$
\\
Exp \cite{pdg}&0.42$\pm$0.02&0.15$\pm$0.02&0.086$\pm$0.042&0.10$\pm$0.02
\\\hline
\end{tabular}
\end{center}
\end{table}

\begin{table}[htb]
\begin{center}
\caption{Ratio $\Gamma(\Upsilon(nS)\to
\ell^+\ell^-)/\Gamma(\Upsilon(1S)\to \ell^+\ell^-)$. `Exp1' are the
experimental data with $\Gamma_{ee}(1S)=1.340\pm0.018$ keV, `Exp2' are the experimental data with $\Gamma_{ee}(1S)=1.29\pm0.09$ keV. For
$\Upsilon(4S)$, $\Gamma_{ee}(4S)=0.272\pm0.029~(0.322\pm0.056)$
keV for the result inside (outside) the brackets.}
\vspace{0.5cm}\label{table7}
\begin{tabular}
{|c|c|c|c|c|}\hline
~~~~~~~~~~~~&~~~$\frac{\Gamma(\Upsilon(2S))}{\Gamma(\Upsilon(1S))}$~~~
&~~~$\frac{\Gamma(\Upsilon(3S))}{\Gamma(\Upsilon(1S))}$~~~
&~~~$\frac{\Gamma(\Upsilon(4S))}{\Gamma(\Upsilon(1S))}$~~~
&~~~$\frac{\Gamma(\Upsilon(5S))}{\Gamma(\Upsilon(1S))}$~~~
\\\hline
Ours&0.488$^{+0.008}_{-0.009}$&0.349$^{+0.003}_{-0.008}$&0.275$^{+0.004}_{-0.000}$&0.229$^{+0.003}_{-0.001}$

\\
Exp1~\cite{pdg}&~~0.457$\pm$0.014~~&0.33$\pm$0.01&~~0.203$\pm$0.024~(0.240$\pm$0.045)~~&0.23$\pm$0.06
\\
Exp2~\cite{pdg}&0.47$\pm$0.04&0.34$\pm$0.03&0.21$\pm$0.04(0.25$\pm$0.06)&0.24$\pm$0.07
\\\hline
\end{tabular}
\end{center}
\end{table}

To cancel the model dependence of the theoretical predictions, we give in Table \ref{table5} and Table \ref{table7} the ratios $\Gamma(\psi(nS)\to e^+e^-)/\Gamma(J/\psi\to e^+e^-)$ and $\Gamma(\Upsilon(nS)\to \ell^+\ell^-)/\Gamma(\Upsilon(1S)\to \ell^+\ell^-)$. For the
$\Upsilon(nS)$ decay, we obtain the same central values for the $e$ and $\tau$ final states, so we only present the ratio
$\Gamma(\Upsilon(nS)\to \ell^+\ell^-)/\Gamma(\Upsilon(1S)\to \ell^+\ell^-)$ in Table \ref{table5} and Table \ref{table7}, which are calculated using the $e^+e^-$ final states listed in Table \ref{table3}.

Table \ref{table5} shows that the ratio $\frac{\Gamma(\psi(2S))}{\Gamma(J/\psi)}$ is larger but close to the experimental data, while the ratios for highly excited states are much larger than the experimental data. Table \ref{table7} shows the bottomonium leptonic decay ratios. In the row `Exp1', the value of $\Gamma_{ee}(1S)=1.340\pm0.018$ keV is used, which is directly listed in PDG. In the row `Exp2', $\Gamma_{ee}(1S)=1.29\pm0.09$ keV is used, which is calculated using the branching ratio of $\Upsilon(1S) \to e^+ e^-$ given in PDG. For $\Upsilon(4S)$, the results outside the brackets were obtained using $\Gamma_{ee}(4S)=0.272\pm0.029$ keV from the PDG, while  the results inside the brackets used $\Gamma_{ee}(4S)=0.322\pm0.056$ keV obtained from the PDG branching ratio. It can be seen that all our theoretical predictions are consistent with the experimental data.

\section{Summary}

In this paper, we studied the leptonic decays of heavy vector quarkonia. For the charmonium decays, not all states are consistent with the experimental data, while for the bottomonium decays, almost all $S$ wave states are in good agreement with the data.

Theoretical results of the ratios $\Gamma(\psi(nS)\to e^+e^-)/\Gamma(J/\psi\to e^+e^-)$ and $\Gamma(\Upsilon(nS)\to \ell^+\ell^-)/\Gamma(\Upsilon(1S)\to \ell^+\ell^-)$ were given in Ref. \cite{radford}, where the potential model was used including the $v_Q^2$ relativistic corrections and pQCD corrections at NLO. These results are comparable with ours, i.e. the charmonium leptonic decay widths are not consistent with the experimental data and the bottomonium leptonic widths are in good agreement with the data. This situation was also observed in Ref. \cite{shah}. It seems that the same theoretical tool cannot provide satisfactory results for both the charmonium and bottomonium systems \cite{rai}. {{There are several possible reasons for this difference in our study. It may be that the instantaneous approximation works well for bottomonium, but is not good enough for charmonium. An improvement of the Cornell potential may be needed, and more importantly, the perturbative QCD corrections may have larger effect in charmonium decays than in bottomonium decays. Since the BS equation is an integral equation, the QCD corrections from the gluon ladder diagrams are already included, but other QCD corrections in the kernel or in the quark propagators may need to be improved in future calculations}}.

The Bethe-Salpeter method provides a strict way to deal with the relativistic effects. In this framework, we found that the relativistic corrections are large and important for the leptonic decays $\psi(nS) \to \ell^+\ell^-$ and $\Upsilon(nS) \to \ell^+\ell^-$. For the $\psi(1S-5S)$ leptonic decays, the relativistic effects are $22^{+3}_{-2}\%$, $34^{+5}_{-5}\%$, $41^{+6}_{-6}\%$, $52^{+11}_{-13}\%$ and $62^{+14}_{-12}\%$, respectively. Therefore, for the highly excited states $\psi(nS)$, the relativistic corrections give dominant contributions. For the $\Upsilon(1S-5S)$ decays, the relativistic effects are $14^{+1}_{-2}\%$, $23^{+0}_{-3}\%$, $20^{+8}_{-2}\%$, $21^{+6}_{-7}\%$ and $28^{+2}_{-7}\%$, respectively. Thus, relativistic effects should be considered for a sound prediction of the heavy quarkonium decays.

\section{Acknowledgments}

This work was supported in part by the National Natural Science Foundation of China (NSFC) under Grant No.~11575048, No.~11625520 and No.~11847301.

\end{document}